\documentclass[12pt]{article}
\usepackage{epsf}

\newcommand{\Lm}{\Lambda}

\newcommand{\Acl}{A_{\rm cl}}

\newcommand{\am}{{\rm am}}
\newcommand{\sn}{{\rm sn}}
\newcommand{\cn}{{\rm cn}}
\newcommand{\dn}{{\rm dn}}
\newcommand{\gef}[1]{g_{{\rm eff}#1}}

\newcommand{\dms}{\Delta m^{2}}

\newcommand{\AR}{A_{\rm R}}
\newcommand{\AI}{A_{\rm I}}
\newcommand{\ar}[1]{a_{{\rm R},#1}}
\newcommand{\ai}[1]{a_{{\rm I},#1}}
\newcommand{\bR}[1]{b_{{\rm R},#1}}
\newcommand{\bI}[1]{b_{{\rm I},#1}}
\newcommand{\mR}[1]{m_{{\rm R},#1}}
\newcommand{\mI}[1]{m_{{\rm I},#1}}
\newcommand{\f}[2]{f^{(#1)}_{#2}}
\newcommand{\ps}[2]{\psi^{(#1)}_{#2}}
\newcommand{\CR}[1]{C_{{\rm R},#1}}

\newcommand{\cL}{{\cal L}}
\newcommand{\del}{\partial}

\newcommand{\dls}{\partial\hspace{-6.5pt}/}

\newcommand\be{\begin{equation}}
\newcommand\ee{\end{equation}}
\newcommand\bea{\begin{eqnarray}}
\newcommand\eea{\end{eqnarray}}

\begin{document}

\thispagestyle{empty}
%
%
%
\begin{center}
{\Large
{\bf SUSY Breaking by Stable non-BPS Walls
 }} 
\vskip 0.5cm

{
 Nobuhito Maru~$^{a}$}
\footnote{\it  e-mail address: 
maru@hep-th.phys.s.u-tokyo.ac.jp},  
{
Norisuke Sakai~$^{b}$}
\footnote{{\it  e-mail address: nsakai@th.phys.titech.ac.jp}, 
Speaker at the Tohwa conference July 3--7,2001},
{
Yutaka Sakamura~$^{b}$}
\footnote{\it  e-mail address: sakamura@th.phys.titech.ac.jp} 
~and~~ {
Ryo Sugisaka~$^{b}$}
\footnote{\it  e-mail address: sugisaka@th.phys.titech.ac.jp}


{ \it   $^{a}$Department of Physics, University of Tokyo 
113-0033, JAPAN \\
$^{b}$Department of Physics, Tokyo Institute of Technology 
\\
Tokyo 152-8551, JAPAN  }
%
%

\end{center}
{\bf Abstract.}
A new simple mechanism for SUSY breaking is proposed due to the 
coexistence of BPS domain walls. 
It is assumed that our world is on a BPS domain wall and that the other 
BPS wall breaks the SUSY preserved by our wall. 
This mechanism requires no messenger fields nor complicated SUSY breaking 
sector on any of the walls. 
We obtain an ${\cal N}=1$ model in four dimensions which 
has an exact solution of a stable non-BPS configuration of two walls. 
We propose that the overlap of 
the wave functions of the N-G fermion and those of physical fields 
provides a practical method to evaluate SUSY breaking mass splitting 
on our wall thanks to a low-energy theorem. 
This is based on our recent works hep-th/0009023 and hep-th/0107204. 
%
\setcounter{page}{1}
\setcounter{footnote}{0}
\renewcommand{\thefootnote}{\arabic{footnote}}

\begin{center}
{\large\bf Introduction}
\end{center}

Recently it is quite popular to consider models where our world is 
embedded in a higher dimensional spacetime as a topological defect
\cite{LED,RS}. 
At the same time, supersymmetry (SUSY) is one of the most promising ideas to 
solve the hierarchy problem in unified theories \cite{DGSW}. 
It has been noted for some years that one of the most 
important issues for 
SUSY unified theories is to understand the SUSY breaking in 
our observable world. 
Domain walls or other topological defects preserving 
part of the original SUSY in the fundamental theory 
are called the BPS states \cite{WittenOlive} in SUSY theories. 
Walls have co-dimension one and typically 
 preserve half of the original SUSY, which are called 
 $1/2$ BPS states  \cite{CGR}. 

The new possibility offered by the brane world 
scenario stimulated  studies of SUSY breaking. 
Recently we have proposed a simple mechanism of SUSY 
breaking due to the 
coexistence of different kinds of BPS domain walls and 
proposed an efficient 
method to evaluate the SUSY breaking parameters such as 
the boson-fermion mass-splitting 
by means of overlap of wave functions involving the 
Nambu-Goldstone (NG) 
fermion \cite{MSSS}.
These points were illustrated by taking a toy model in 
four dimensions, 
which allows an exact solution of coexisting walls with a 
three-dimensional 
effective theory. 
Although the first model was only meta-stable, we were able to 
show approximate evaluation of the overlap allows us to determine 
the mass-splitting reliably. 
More recently, we have constructed a stable non-BPS configuration of 
two walls in an ${\cal N}=1$ supersymmetric model in four 
dimensions to demonstrate our idea of SUSY 
breaking due to 
the coexistence of BPS walls. 
We have also extended our analysis 
to more realistic case of four-dimensional effective theories 
and examined the consequences of our mechanism in detail \cite{MSSS2}. 

Our proposal for a SUSY breaking mechanism requires no 
messenger fields, 
nor  complicated SUSY breaking sector on any of the walls. 
We assume that our world is on a wall and SUSY is broken 
only by the 
coexistence of another wall with some distance from our wall. 
The NG fermion is localized on the distant 
wall and its overlap with the wave functions of physical fields 
on our wall gives the boson-fermion mass-splitting of 
physical fields on our 
wall thanks to a low-energy theorem \cite{lee-wu}. 
We have proposed that this overlap provides a practical 
method to evaluate 
the mass-splitting in models with SUSY breaking due to the 
coexisting walls.

\begin{center}
{\large\bf Stable non-BPS configuration of two walls}
\end{center}
\label{WS:solublemodel}

We introduce a simple four-dimensional Wess-Zumino model 
as follows.
\begin{eqnarray}
 &\!\!\!\cL&\!\!\!=\bar{\Phi}\Phi |_{\theta^{2}\bar{\theta}^{2}}
  +W(\Phi)|_{\theta^{2}}+{\rm h.c.},  \label{Logn} 
\qquad 
W(\Phi)
=\frac{\Lm^{3}}{g^{2}}\sin\left(\frac{g}{\Lm}\Phi\right), 
\end{eqnarray}
where $\Phi=(A, \Psi, F)$ is a chiral superfield. 
A scale parameter $\Lm$ has a mass-dimension one and a 
coupling constant 
$g$ is dimensionless, and both of them are real positive.
We choose $y=X^{2}$ as the extra dimension 
and compactify it on $S^{1}$ of radius $R$.
Other coordinates are denoted as $x^{m}$ ($m=0,1,3$). 
The bosonic part of the model is 
\be
 \cL_{\rm 
bosonic}=-\del^{\mu}A^{\ast}\del_{\mu}A-\frac{\Lm^{4}}{g^{2}}
 \left|\cos\left(\frac{g}{\Lm}A\right)\right|^{2}.
\ee
The target space of the scalar field $A$ has a topology of a 
cylinder. 
This model has two vacua at $A=\pm\pi\Lm/(2g)$, both lie 
on the real axis.

Let us first consider the case of the limit $R\to\infty$.
In this case, there are two kinds of BPS domain walls in this 
model.
One of them is 
\be
 \Acl^{(1)}(y)=\frac{\Lm}{g}\left\{2\tan^{-1}e^{\Lm(y-y_{1})}
 -\frac{\pi}{2}
 \right\},
 \label{eq:first_wall}
\ee
which interpolates the vacuum at $A=-\pi\Lm/(2g)$ to that 
at 
$A=\pi\Lm/(2g)$ as $y$ increases from $y=-\infty$ to 
$y=\infty$.
The other wall is 
\be
 \Acl^{(2)}(y)=\frac{\Lm}{g}\left\{-2\tan^{-1}e^{-\Lm(y-y_{2})}+
\frac{3\pi}{2}
 \right\},
 \label{eq:second_wall}
\ee
which interpolates the vacuum at $A=\pi\Lm/(2g)$ to that at 
$A=3\pi\Lm/(2g)=-\pi\Lm/(2g)$.
Here $y_{1}$ and $y_{2}$ are integration constants and 
represent 
the location of the walls along the extra dimension. 
The four-dimensional supercharge $Q_{\alpha}$ can be 
decomposed into 
two two-component Majorana supercharges $Q^{(1)}_{\alpha}$ 
and 
$Q^{(2)}_{\alpha}$ which can be regarded as supercharges in 
three dimensions 
\be
 Q_{\alpha}=\frac{1}{\sqrt{2}}(Q^{(1)}_{\alpha}+iQ^{(2)}_{\alpha}).
\ee
Each wall breaks a half of the bulk supersymmetry: 
 $Q^{(1)}_{\alpha}$ is broken by $\Acl^{(2)}(y)$, and $Q^{(2)}_{\alpha}$ 
 by $\Acl^{(1)}(y)$.
Thus all of the bulk supersymmetry will be broken if these 
walls coexist.

We will consider such a two-wall system to study the SUSY 
breaking 
effects in the low-energy three-dimensional theory on the 
background. 
The field configuration of the two walls will wrap around the 
cylinder in 
the target space of $A$ as $y$ increases from $0$ to $2\pi 
R$. 
Such a configuration should be a solution of the equation of 
motion, 
\be
 \del^{\mu}\del_{\mu}A+\frac{\Lm^{3}}{g}\sin\left(\frac{g}{\Lm}
A^{\ast}\right)
 \cos\left(\frac{g}{\Lm}A\right)=0. \label{EOM1}
\ee
We find that a general real static solution of Eq.(\ref{EOM1}) 
that depends 
only on $y$ is 
\be
 \Acl(y)=\frac{\Lm}{g}\am\left(\frac{\Lm}{k}(y-y_{0}),k\right), 
 \label{A_classical}
\ee
where $k$ and $y_{0}$ are real parameters and the function 
$\phi=\am(u,k)$ 
denotes the amplitude function, which is defined 
as an inverse function of 
\be
 u(\varphi)=\int_{0}^{\varphi}\frac{{\rm d}\theta}
 {\sqrt{1-k^{2}\sin^{2}\theta}}.
\ee
If $k>1$, it becomes a periodic function with the period 
$4K(1/k)/\Lm$, where the function $K(k)$ is the complete 
elliptic integral 
of the first kind.
If $k<1$, the solution $\Acl(y)$ is a monotonically increasing 
function with 
\be
 \Acl\left(y+{4kK(k) \over \Lambda}\right)= \Acl(y)+ 2 
\pi{\Lambda \over g}.
\ee
Since the field $A$ is an angular variable 
$A=A+2\pi\Lambda/g$, 
we can choose the compactified radius 
$2\pi R=4kK(k)/\Lm$ so that the classical field configuration 
$\Acl(y)$ contains two walls and becomes 
periodic modulo $2\pi\Lambda/g$. 
We shall take $y_{0}=0$ to locate one of the walls at $y=0$. 
Then we find that the other wall is located at the anti-podal 
point 
 $y=\pi R$ of the compactified circle. 

In the limit of $R\to\infty$, {\em i.e.}, $k\to 1$, $\Acl(y)$ 
approaches to 
the BPS configuration $\Acl^{(1)}(y)$ with $y_1=0$ near $y=0$, which 
preserves $Q^{(1)}$, 
and to $\Acl^{(2)}(y)$ with $y_2=\pi R$ near $y=\pi R$, which preserves 
$Q^{(2)}$. 
We will refer to the wall at $y=0$ as ``our wall'' and the wall at 
$y=\pi R$ as ``the other wall''.

\begin{center}
{\large\bf The fluctuation mode expansion}
\end{center}
Let us consider the fluctuation fields around the background 
$\Acl(y)$, 
\bea
 A(X)&\!\!=&\!\!\Acl(y)+\frac{1}{\sqrt{2}}(\AR(X)+i\AI(X)), 
 \nonumber\\
 \Psi_{\alpha}(X)&\!\!=&\!\!\frac{1}{\sqrt{2}}(\Psi_{\alpha}^{(1)}(
X)
 +i\Psi_{\alpha}^{(2)}(X)). \label{fluc_fields}
\eea
To expand them in modes, we define 
the mode functions as solutions of equations:
\bea
 \left\{-\del_{y}^{2}-\Lm^{2}\cos\left(\frac{2g}{\Lm}\Acl(y) 
 \right)\right\}
 \bR{p}(y)&\!\!=&\!\!\mR{p}^{2}\bR{p}(y), \nonumber\\
 \{-\del_{y}^{2}+\Lm^{2}\}\bI{p}(y)&\!\!=&\!\!\mI{p}^{2}(y)\bI{p}(y
), 
 \label{boson_mode_eq}
\eea
\bea
 \left\{-\del_{y}-\Lm\sin\left(\frac{g}{\Lm}\Acl(y)\right)\right
\}
 \f{1}{p}(y)&\!\!=&\!\!m_{p}\f{2}{p}(y), \nonumber\\
 \left\{\del_{y}-\Lm\sin\left(\frac{g}{\Lm}\Acl(y) 
 \right) \right\}
 \f{2}{p}(y)&\!\!=&\!\!m_{p}\f{1}{p}(y). \label{fermion_mode_eq}
\eea
The four-dimensional fluctuation fields can be expanded as 
\be
 \AR(X)=\sum_{p}\bR{p}(y)\ar{p}(x),\;\;\;
 \AI(X)=\sum_{p}\bI{p}(y)\ai{p}(x), 
\label{eq:boson_mode_decomp}
\ee
\be
 \Psi^{(1)}(X)=\sum_{p}\f{1}{p}(y)\ps{1}{p}(x),\;\;\;
 \Psi^{(2)}(X)=\sum_{p}\f{2}{p}(y)\ps{2}{p}(x). 
 \label{eq:fermion_mode_decomp}
\ee
As a consequence of the linearized equation of motion, the 
coefficient 
$\ar{p}(x)$ and $\ai{p}(x)$ are scalar fields in 
three-dimensional effective theory with masses $\mR{p}$ and 
$\mI{p}$,
and $\ps{1}{p}(x)$ and $\ps{2}{p}(x)$ are three-dimensional 
spinor fields with masses $m_{p}$, respectively.

Exact mode functions and mass-eigenvalues are known for 
several light modes 
of $\bR{p}(y)$, 
\bea
 \bR{0}(y)&\!\!=&\!\!\CR{0}\dn\left(\frac{\Lm y}{k},k\right), 
 \;\;\; \mR{0}^{2}=0, \nonumber \\
 \bR{1}(y)&\!\!=&\!\!\CR{1}\cn\left(\frac{\Lm y}{k},k\right), \;
\;\; 
 \mR{1}^{2}=\frac{1-k^{2}}{k^{2}}\Lm^{2}, \nonumber \\
 \bR{2}(y)&\!\!=&\!\!\CR{2}\sn\left(\frac{\Lm y}{k},k\right), \;
\;\; 
 \mR{2}^{2}=\frac{\Lm^{2}}{k^{2}}, \label{bosonR_mode_fnc}
\eea
where functions $\dn(u,k)$, $\cn(u,k)$, $\sn(u,k)$ are the 
Jacobi's 
elliptic functions and $\CR{p}$ are normalization factors. 
{}For $\bI{p}(y)$, we can find all the eigenmodes 
\be
 \bI{p}(y)=\frac{1}{\sqrt{2\pi R}}{\rm e}^{i{p \over R}y}
,\;\;\; 
 \mI{p}^{2}=\Lm^{2}+\frac{p^{2}}{R^{2}},\;\;\;
 ( p \in {\bf Z} 
 ).
\ee
The massless field $\ar{0}(x)$ is the Nambu-Goldstone (NG) 
boson 
for the breaking of the translational invariance in the extra 
dimension.
The first massive field $\ar{1}(x)$ corresponds to the 
oscillation of the 
background wall around the anti-podal equilibrium point and 
hence becomes 
massless in the limit of $R\rightarrow \infty$. 
All the other bosonic fields remain massive in that limit. 

{}For fermions, only zero modes are known explicitly,
\be
 \f{1}{0}(y)=C_{0}\left\{\dn\left(\frac{\Lm y}{k},k\right)
 +k\cn\left(\frac{\Lm y}{k},k\right)\right\}, \;\;\;
 \f{2}{0}(y)=C_{0}\left\{\dn\left(\frac{\Lm y}{k},k\right)
 -k\cn\left(\frac{\Lm y}{k},k\right)\right\}, 
\label{fermion_mode_fnc}
\ee
where $C_{0}$ is a normalization factor.
These fermionic zero modes are the NG fermions for the 
breaking of 
$Q^{(1)}$-SUSY and $Q^{(2)}$-SUSY, respectively. 

Thus there are four fields which are massless or become 
massless in the limit 
of $R\rightarrow \infty$: 
$\ar{0}(x)$, $\ar{1}(x)$, $\ps{1}{0}(x)$ and $\ps{2}{0}(x)$.
Other fields are heavier and have masses of the order of 
$\Lm$. 

In the following discussion, we will concentrate ourselves on 
the breaking of the $Q^{(1)}$-SUSY, which is approximately 
preserved 
by our wall at $y=0$.
So we call the field $\ps{2}{0}(x)$ the NG fermion 
in the rest of the paper.

\begin{center}
{\large\bf Three-dimensional effective Lagrangian}
\end{center}
We can obtain a three-dimensional effective Lagrangian 
by substituting the mode-expanded fields 
Eq.(\ref{eq:boson_mode_decomp}) 
and Eq.(\ref{eq:fermion_mode_decomp}) into the Lagrangian 
(\ref{Logn}), 
and carrying out an integration over $y$ 
\begin{eqnarray}
 \cL^{(3)}&\!\!\!=&\!\!\!
 -V_{0}-\frac{1}{2}\del^{m}\ar{0}\del_{m}\ar{0}
 -\frac{1}{2}\del^{m}\ar{1}\del_{m}\ar{1}-\frac{i}{2}\ps{1}{0}
 \dls\ps{1}{0}
 -\frac{i}{2}\ps{2}{0}\dls\ps{2}{0} \nonumber \\
 &\!\!\!&\!\!\!
 -\frac{1}{2}\mR{1}^{2}\ar{1}^{2}+\gef{}\ar{1}\ps{1}{0}\ps{2}{0}
 +\cdots, 
 \label{effthry}
\end{eqnarray}
where $\dls\equiv\gamma^{m}_{(3)}\del_{m}$ and 
an abbreviation denotes terms involving heavier fields and 
higher-dimensional terms. 
Here 
 $\gamma$-matrices in three dimensions are defined by 
 $\left(\gamma^{m}_{(3)}\right)
 \equiv\left(-\sigma^2, i\sigma^3, -i\sigma^1\right)$. 
The vacuum energy $V_0$ is given by the energy density of 
the background 
and thus 
\begin{equation}
 V_{0}\equiv \int^{\pi R}_{-\pi R}{\rm 
d}y\left\{\left(\del_{y}\Acl\right)
 +\frac{\Lm^{4}}{g^{2}}\cos^{2}\left(\frac{g}{\Lm}\Acl 
 \right) \right\}
 =\frac{\Lm^{3}}{g^{2}k}\int_{-2K(k)}^{2K(k)}{\rm d}u\left\{
 (1+k^{2})-2k^{2}\sn^{2}(u,k)\right\},
\label{eq:vacuum_energy}
\end{equation}
and the effective Yukawa coupling $\gef{}$ is 
\begin{equation}
 \gef{}\equiv\frac{g}{\sqrt{2}}\int^{\pi R}_{-\pi R}{\rm d}y \, 
 \cos\left(\frac{g}{\Lm}\Acl(y)\right) \bR{1}(y)\f{1}{0}(y)\f{2}{0}
(y)
 =\frac{g}{\sqrt{2}}\frac{C_{0}^{2}}{\CR{1}}(1-k^{2}).
 \label{geff}
\end{equation}

In the limit of $R\to\infty$, the parameters $\mR{1}$ and 
$\gef{}$ vanish 
and thus we can redefine the bosonic massless fields 
as 
\be
 \left(\begin{array}{c}a^{(1)}_{0} \\ a^{(2)}_{0}\end{array}\right)
 =\frac{1}{\sqrt{2}}\left(\begin{array}{cc} 1 & 1 \\ -1 & 1 
\end{array}\right)
 \left(\begin{array}{c}\ar{0} \\ \ar{1}\end{array}\right). 
 \label{boson-mixing}
\ee
In this case, the fields $a^{(1)}_{0}(x)$ and $\ps{1}{0}(x)$ 
form a supermultiplet 
for $Q^{(1)}$-SUSY and their mode functions are both 
localized on our wall. 
The fields $a^{(2)}_{0}(x)$ and $\ps{2}{0}(x)$ are singlets 
for $Q^{(1)}$-SUSY and are localized on the other 
wall.\footnote{
The modes $a^{(2)}_{0}(x)$ and $\ps{2}{0}(x)$ form a 
supermultiplet for 
$Q^{(2)}$-SUSY.
} 

When the distance between the walls $\pi R$ is finite, 
$Q^{(1)}$-SUSY is broken and the mass-splittings between 
bosonic and fermionic 
modes are induced.
The mass squared $\mR{1}^{2}$ in Eq.(\ref{effthry}) 
corresponds to 
the difference of the mass squared $\dms$ 
between $a^{(1)}_{0}(x)$ and $\ps{1}{0}(x)$ 
since the fermionic mode $\ps{1}{0}(x)$ is massless.
Besides the mass terms, we can read off the SUSY breaking 
effects 
from the Yukawa couplings like $\gef{}$.

We have noticed in Ref.\cite{MSSS} that these two SUSY 
breaking parameters, 
$\mR{1}$ and $\gef{}$, 
are related by the low-energy theorem associated with the 
spontaneous breaking of SUSY. 
In our case, the low-energy theorem becomes 
\begin{equation}
 \frac{\gef{}}{\mR{1}^{2}}=\frac{1}{2f}. \label{GTR-gef}
\end{equation}
where $f$ is an order parameter of the SUSY breaking, and it 
is given 
by the square root of the vacuum (classical background) 
energy density $V_0$ in Eq.(\ref{eq:vacuum_energy}). 
Since the superpartner of the fermionic field $\ps{1}{0}(x)$ 
is a mixture of mass-eigenstates, 
we had to take into account the mixing 
Eq.(\ref{boson-mixing}).

The mass-splitting decays exponentially 
as the wall distance increases.
This is one of the characteristic features of our SUSY 
breaking mechanism. 
This fact can be easily understood by remembering the profile 
of each modes. 
Note that the mass-splitting $\dms(=\mR{1}^{2})$ is 
proportional 
to the effective Yukawa coupling constant $\gef{}$, 
which is represented by an overlap integral of the mode 
functions. 
Here the mode functions of the fermionic field $\ps{1}{0}(x)$ 
and 
its superpartner are both localized 
on our wall, and that of the NG fermion $\ps{2}{0}(x)$ is 
localized 
on the other wall.
Therefore the mass-splitting becomes exponentially small 
when the distance 
between the walls increases, because of exponentially 
dumping tails 
of the mode functions.

We have worked out in Ref.~\cite{MSSS2} 
how various soft SUSY breaking terms 
can arise in our framework. 
Phenomenological implications have also been briefly discussed.


This work is supported in part by Grant-in-Aid for Scientific 
Research from the Ministry of Education, Culture, Sports, 
Science and 
Technology,Japan, priority area(\#707) ``Supersymmetry and 
unified theory
of elementary particles" and No.13640269. 
N.M.,Y.S.~and R.S.~are supported 
by the Japan Society for the Promotion of Science for Young 
Scientists 
(No.08557, No.10113 and No.6665).


\end{document}